\documentstyle[12pt]{article}
\def \be{\begin{equation}}
\def \ee{\end{equation}}
\def \lb{\label}
\def \ba{\begin{array}{l}}
\def \ea{\end{array}}

\def \t{\tau}
\def \a{\alpha}

\def \d{\delta}

\def \e{\epsilon}
\def \f{\phi}

\def \lm{\lambda}

\def \tl{\tilde}
\def \ol{\overline}

\def \fr{\frac}

\def \2{\frac{1}{2}}
\def \4{\frac{1}{4}}
\begin{document}

\begin{center}

{\Large \bf Stability of the Renormalization Group in the 2D Random Ising
and Baxter Models with respect to the Replica Symmetry Breaking}

\vskip .4in

D.E.Feldman, A.V.Izyumov and Viktor Dotsenko

\vskip .2in

Landau Institute for Theoretical Physics,\\
Russian Academy of Sciences, \\
Kosygina 2, 117940 Moscow, Russia\\

\end{center}

\begin{abstract}
We study the critical properties of the weakly disordered two-dimensional
Ising and Baxter models in terms of the renormalization group (RG) theory
generalized to take into account the replica symmetry breaking (RSB) effects.
Recently it has been shown that the traditional replica-symmetric RG flows
in the dimensions $D = 4 - \e$ are unstable with respect to
the RSB potentials  and a new spin-glass type critical phenomena
has been discovered \cite{dhss},\cite{df}. In contrast, here it is demonstrated
that in the considered two-dimensional systems the renormalization-group flows
are {\it stable} with respect to the RSB modes. It is shown that the
solution of the renormalization group equations with arbitrary starting
RSB coupling matrix exhibits asymptotic approach to the traditional
replica-symmetric ones. Thus, in the leading order the non-perturbative
RSB degrees of freedom does not effect the critical phenomena in the
two-dimensional weakly disordered Ising and Baxter models studied earlier.
\end{abstract}

\newpage

The effects produced by weak quenched disorder on the
critical phenomena near the phase transition point have been studied since many
years ago \cite{harris}-\cite{numer}.
According to the Harris criterion \cite{harris}, the disorder effects
the critical behaviour only if $\alpha$, the specific heat exponent of the
pure system, is positive. In this case a new universal critical behavior, with
new critical exponents, is established sufficiently close to the phase
transition point for $ (T/T_{c} - 1) \equiv \t \ll \t_{u} \equiv u^{1/\a}$
\cite{new}, where $u \ll 1$ is the parameter which describes the strength
of the disorder. In contrast, when $\a < 0$, the disorder appears to be
irrelevant for the critical behavior.

Originally the modified critical behaviour
has been derived for the classical $\f^{4}$ model near four dimensions
\cite{new}, and later it has been studied for the two-dimensional
Ising \cite{ising}, Baxter \cite{baxter} and Potts \cite{potts}
models by various renormalization group (RG) techniques, and by numerical
simulations \cite{numer}.

In dealing with the quenched disorder the
traditional approach is the replica method, and in terms of replicas all the
results obtained for the systems listed above correspond to the so-called
replica-symmetric (RS) solutions. Physically it means that olny unique
ground state is assumed to be relevant for the observable thermodynamics.
The problem, however, is that in the presence of the quenched disorder there
exist numerous local minimum configurations separated by finite barriers, and
in this case the direct application of the traditional replica-symmetric RG
scheme may be questioned.

On the other hand, it is the Parisi Replica Symmetry Breaking (RSB) scheme
which has been developed specifically for dealing with disordered systems
which exhibit numerous local minima states (see e.g. \cite{sg}). Recent
studies show that besides mean-field theory of spin-glasses the RSB approach
can also be generalized for situations where one has to deal with
fluctuations as well \cite{manifold},\cite{rsb-Marc},\cite{Korshunov}.
In the paper \cite{dhss} qualitative arguments were presented demonstrating the
mechanism how the summation over multiple local minima configurations could
provide additional non-trivial RSB interaction
potentials for the fluctuating fields.

One can hope that such type of the generalized RG scheme self-consistently
takes into account relevant degrees of freedom coming from the
numerous local minima, and in particular, the instability of the RS fixed point
with respect to the RSB would indicate that the multiplicity of the local
minima is relevant for the critical properties in the fluctuation region.
And inversely, if the traditional RS RG flows turn out to be stable with
respect to the RSB modes, then it can be concluded that such non-perturbative
degrees of freedom are irrelevant for the critical behaviour in a considered
system.

The first example of replica symmetry broken solutions in the
renormalization group has been suggested in  \cite{doussal} in the context of
the 2D random field XY-model, where instability of the RG flows
with respect to the RSB modes has been discovered.
Similar phenomena has been observed in the weakly disordered classical
ferromagnet (the $\f^{4}$-theory) in the dimension $D = 4 - \e$ \cite{dhss}.
Here the RSB degrees of freedom produce dramatic effect on the asymptotic
behaviour of the RG flows, such that for a general type of the RSB there
exist no stable fixed points, and the RG arrives into the strong coupling
regime at the finite scale ($\sim \exp(-1/u)$) \cite{df}.

On the other hand, similar considerations for weakly disordered 2D
Potts model shows that, although the traditional RS fixed point also turns
out to be unstable, there exist stable non-trivial fixed point
characterized by the continuous RSB structure of the coupling matrix
\cite{rsb-potts}.

In this Letter we report on the RSB solution for the two-dimensional Ising and
Baxter models with random bonds. We consider the models
with spin couplings having small fluctuations around a mean ferromagnetic
value. This gives a possibility to study the model in the continuum limit,
because one approaches the critical point sufficiently close before the
randomness becomes relevant. For the two-dimensional Ising and Baxter models
this allows to use the renormalization group based
on the fermion representation.

In contrast to the previous studies, in the considered systems the RG flows
turns out to be stable with respect to RSB modes. The explicit solution of the
corresponding RG equations shows that for any starting RSB structure
of the coupling matrix the asymptotic solutions become replica-symmetric.
It means that the RSB ("non-perturbative") degrees of freedom appear to be
irrelevant for the critical phenomena in these systems, and they exhibit the
usual replica symmetric critical behavior which has been studied earlier
\cite{ising},\cite{baxter}.

\vspace{5mm}

{\bf The Ising Model}

It is well known that the two-dimensional Ising model in the critical
region is equivalent to the problem of the free fermions \cite{HG}.
The effect of the quenched disorder can be described by the random
contribution to the effective temperature $\tau$. Thus, the random bond
Ising model in the critical region can be reduced to the field theory with
the following Lagrangian:

\be
L[\psi] = -\frac{1}{2}\int d^2x[\bar\psi\hat\partial\psi+
(\tau+\delta\tau)\bar\psi\psi],
\ee
where $\delta\tau$ is the Gaussian random variable with
$\ol{(\d\t^{2})} = 2u \ll 1$, and $\psi$ and $\bar\psi=i\psi\hat\sigma_y$ are
the two-component real fermion fields.

The averaging over quenched random variable $\delta\tau$ can be performed in
terms of the standard replica procedure. The resulting
replica Lagrangian has the following form:

\be
L_{n}[\psi]=-\int
d^2x[\frac{1}{2}\sum_{a=1}^{n}\bar\psi^a(\hat\partial+\tau)\psi^a-
\frac{1}{4}\sum_{a,b=1}^{n}g_{ab}\bar\psi^a\psi^a\bar\psi^b\psi^b],
\ee
In the usual replica-symmetric theory the coupling matrix does not
depend on the replica indices: $g_{ab} = u$. Then, in terms of the
standard RG scheme one can easily calculate the critical properties
of the system, and in particular for leading singularity of the specific heat
one finds: $C(\tau)\sim\log\log(1/|\tau|)$ \cite{ising}.

According to the discussion in the introductory part of this Letter,
in our present scenario we assume that due to non-perturbative
effects the replica symmetry in the coupling matrix $g_{ab}$ can be broken.
In other words, the starting point of the further analysis is the
assumption that the matrix $g_{ab}$ has the Parisi-type structure
(see e.g. \cite{sg}).

The corresponding (one-loop) RG equations for the replica matrix $g_{ab}$
and for the mass parameter $\t$ are:

\be
\label{g1}
\frac{dg_{ab}}{d\xi}=\frac{1}{\pi}\sum_{c\not=a,b}
g_{ac}g_{cb}.
\ee
\be
\label{t1}
\frac{d\log\tau}{d\xi}=\frac{1}{\pi}\sum_{b\not=1}g_{1b}.
\ee
where $\xi$ is the standard RG rescaling parameter.

In terms of the Parisi RSB scheme \cite{sg} in the limit $n \to 0$ the
matrix $g_{a\not= b}$ is parametrized by the function $g(x)$ defined in the
interval $0 \leq x \leq 1$ (replica-symmetric situation corresponds to
$g(x) = const$, independent of $x$). According to the standard technique of the
Parisi RSB algebra \cite{rsb-Marc} (see also \cite{dhss},\cite{df})
the equation (\ref{g1}) can be represented as follows:

\be
\label{g2}
\frac{dg(x)}{d\xi}=-\frac{1}{\pi}[xg^{2}(x)+2g(x)\int_x^1 dy g(y)+
\int_0^x dy g^2(y)],
\ee

The structure of this equation is similar to that for the $\phi^4$-theory with
the broken replica symmetry, and it can be solved following
the method suggested in \cite{df}. Simple calculations yield:

\be
\label{g3}
g'(x,\xi)= \frac{g'(x,0)\exp(-\frac{2}{\pi}\int_0^\xi d\eta \bar g(\eta))}
{[1+\fr{1}{\pi}\int_0^x yg'(y,0) dy \int_0^\xi d\eta
\exp(-\frac{2}{\pi}\int_0^\eta d\theta \bar g(\theta))]^2}.
\ee
\be
\label{g4}
g(0,\xi)=g(0,0)\exp(-\frac{2}{\pi}\int_0^\xi\bar g(\eta)d\eta).
\ee
where $g'(x,\xi) \equiv \fr{d}{dx}g(x,\xi)$ and
$\bar g(\xi) \equiv \int_0^1 g(x,\xi)dx$.

Equations (\ref{g3})-(\ref{g4}) allow to study the large scale asymptotic
behaviour of the renormalized function $g(x,\xi)$ provided the starting
function is given. Simple calculations show that at
any value of $x$ in the limit $\xi\to\infty$ the derivative in (\ref{g3})
tends to zero as $[\xi(\log\xi)^{2}]^{-1}$, while
$\bar g(\xi) \sim \xi^{-1}$. It means that the asymptotic renormalized
function $g(x,\xi\to\infty)$ tends to become flat in the whole interval
$0 < x < 1$). In other words, for any starting RSB function
$g(x,\xi=0)$ the asymptotic solution of the RG eq.(\ref{g2}) becomes
replica-symmetric.

One can also easily check that slowly decaying corrections to the
RS solution does not effect the leading singularity of the specific heat.
{}From the eq.(\ref{t1}) for the renormalized temperature parameter
one finds:

\be
\label{t2}
\tau(\xi)=\tau(0)\exp(-\frac{1}{\pi}\int_0^\xi \bar g(\eta) d\eta).
\ee
Then, using eqs. (\ref{g3}),(\ref{g4}) and (\ref{t2}) one can derive the
following closed form relation on $\tau(\xi)$:
\be
\label{t3}
\log(\frac{\tau(0)}{\tau(\xi)})=\frac{g(0,0)}{\pi}\int_0^\xi d\eta
(\frac{\tau(\eta)}{\tau(0)})^2+O(\sqrt{\int_0^\xi d\eta
(\frac{\tau(\eta)}{\tau(0)}})^2).
\ee
In the asymptotic limit $\xi\to\infty$ one finally gets:

\be
\label{t4}
\tau(\xi)\sim\frac{1}{\sqrt{\xi}}+O(\frac{1}{\sqrt{\xi\log\xi}}).
\ee
which in the leading order coincides with the  RS result.
Correspondingly, the leading singularity of the specific heat
(which can be estimated as $C(\t) \sim \int^{\ln(1/|\t|)}d\xi \t^{2}(\xi)$)
remains the same as in the RS case \cite{ising} as well.

\vspace{5mm}

{\bf The Baxter Model}

The Baxter model \cite{B} can be formulated in terms of two 2D Ising models
coupled by four-spin interactions. The strength of this coupling is described
by parameter $\lambda$ (the case of $\lambda =0$ corresponds to two
independent 2D Ising models).
The pure Baxter model can be solved exactly and in particular, its specific
heat exponent is proportional to $\lambda$:
$C_{pure}(\tau ) \sim {|\tau |}^{-\frac{\lambda}{\pi}}$ (for $|\lambda |\ll
1$).

Similar to the weakly disordered 2D Ising system considered above,
in the continuum limit near the critical point the corresponding Baxter model
can be described in terms of the two-component complex fermion field:
\be
\lb{f3}
L[\psi] = -\int d^2 x \lbrack \frac{1}{2} \bar \psi \hat \partial
\psi +\frac{1}{2} (\tau + \delta \tau (x)) \bar \psi \psi
- \4\lambda (\bar \psi \psi)(\bar \psi \psi) \rbrack
\ee
After averaging over the random function $\d\t(x)$ one gets the following
replica Lagrangian:

\be
\label{f8}
L_{n}[\psi]  = -\int d^2x\lbrack
\2\sum_{a=1}^{n} \bar\psi^a (\hat\partial +\t)\psi^a
- \4\sum_{a,b}g_{ab}(\bar\psi^a \psi^a)(\bar\psi^b \psi^b)\rbrack
\ee

In the RS case for the replica coupling matrix one has:
$g_{ab} = \lambda \d_{ab} + u$. The critical properties of such system
has been studied earlier \cite{baxter}, and it was shown that for
$\lambda >0$ (when the specific heat of the pure system is divergent)
the leading singularity of the specific heat coincides with that
of the random Ising model:
$C(\tau )\sim \ln \ln \frac{1}{|\tau |}$.
On the other hand, for $\lambda <0$ (when the specific heat of the pure system
is finite), the singularity of the specific heat changes to:
$C(\tau ) \sim |\tau |^{-\fr{\lm_{*}}{\pi}}$, where
$\lm_{*} = \lm \exp(-u/|\lm|)$.

Here we study the situation when the coupling matrix $g_{ab}$ has a general
RSB Parisi structure. The corresponding RG equations are:

\be
\label{f9}
\frac{dg_{ab}}{d\xi} = -\frac{1}{\pi}\lbrace g_{ab} (g_{aa}+g_{bb})
-2\sum_c g_{ac} g_{cb} \rbrace
\ee
\be
\label{f10}
\frac{d(\ln \tau)}{d\xi} = -\frac{1}{\pi} \lbrace g_{aa}-2\sum_c g_{ac}
\rbrace
\ee

According to the standard technique of the
Parisi RSB algebra, in the limit $n \to 0$ the matrix $g_{ab}$ is
parametrized in terms of its diagonal element $\tilde g$ and the
off-diagonal function $g(x)$. The replica symmetric situation
corresponds to the case $g(x)=const$ independent of $x$.
After simple algebra, instead of the eqs.(\ref{f9}),(\ref{f10})
in the limit $n \to 0$ one gets:

\be
\label{f11}
\ba
\frac{d}{d\xi}\tl g =  -\fr{2}{\pi}\int_0^1 dx g^2(x) \\
\\
\frac{d}{d\xi}g(x) =  \fr{2}{\pi} \lbrack \tl g g(x) - 2\bar g g(x) -\int_0^x
dy
[g(y)-g(x)]^2 \rbrack \\
\\
\frac{d}{d\xi}(\ln\t) = \fr{1}{\pi}(\tl g - 2\bar g)
\ea
\ee
where $\bar g \equiv \int_0^1 dx g(x)$.

The solution of the eqs.(\ref{f11}) can be represented in the same way as
in the Ising case considered above (cf eq.(\ref{g3})):

\be
\lb{f14}
g'(x,\xi)= \frac{g'(x,0)
\exp\{-\frac{2}{\pi}\int_0^\xi d\eta (2\bar g(\eta) - \tl g(\eta))\} }
{[1+\fr{2}{\pi}\int_0^x y g'(y,0) dy \int_0^\xi d\eta
\exp\{-\frac{2}{\pi}\int_0^\eta d\theta (2\bar g(\theta)-\tl g(\theta))\}]^2}.
\ee

Using the above equation one can easily study the asymptotic
(for $\xi \to \infty$) behaviour of the function $g(x,\xi)$.
The behaviour of the solution depends
on the sign of the starting parameter
$g_0 \equiv \tilde g(\xi=0)- \bar g(\xi=0)$ which
corresponds to the initial coupling $\lm$ in the replica symmetric case.
Consider the two cases separately.

1) $g_0 > 0$ \\
The corresponding replica-symmetric RG equations have the following asymptotic
solution \cite{baxter}:
\be
\label{f13}
\ba
g_{RS} \simeq \fr{2}{\pi}[\frac{1}{\xi} - \fr{1}{\xi\ln\xi}]
+ \fr{a}{\xi(\ln\xi)^{2}}\\
\\
\tl g_{RS} \simeq \fr{2}{\pi}[\frac{1}{\xi} - \fr{2}{\xi\ln\xi}]
+ \fr{b}{\xi(\ln\xi)^{2}}
\ea
\ee
where $a$ and $b$ are constants. Then, using eq.(\ref{f14}) for the RSB
correction to these solutions: $g(x,\xi) = g_{RS}(\xi) + g_{RSB}(x,\xi)$ and
$\tl g(\xi) = \tl g_{RS}(\xi) + \tl g_{RSB}(\xi)$, one immediately
finds that

\be
\ba
g'_{RSB}(x,\xi) \sim \fr{V(x)}{\xi(\ln\xi)^{2}}\\
\\
\tl g_{RSB}(\xi) \sim \fr{1}{\xi^{2}(\ln\xi)^{2}}
\ea
\ee
where $V(x)$ is a function which is defined by the starting RSB function
$g(x,\xi=0)$. This means that on large scales the RSB deviations are vanishing
in comparison with the RS solution, and the asymptotical properties
of the model are not affected by the starting RSB.

2) $g_0 < 0 $ \\
In this case the replica-symmetric RG trajectories arrive
to the fixed point $(g_{*}, \tl g_{*})$, which depends on the initial
coupling parameters $g(\xi=0)\equiv g(0)$ and $g_{0}$ \cite{baxter}:

\begin{equation}
\label{f18}
\ba
\tilde g_{*} = - |g_0|\exp\left[ -\frac{g(0)}{|g_0|}\right]\\
\\
g_{*} = 0
\ea
\end{equation}
Using eqs.(\ref{f11}) or (\ref{f14}) this fixed point can be easily shown to
be also stable with respect to general RSB deviations. In other words, the RG
trajectories defined by eqs.(\ref{f11}) arrive to the same
fixed point $(\tl g_{*} ,0)$ as in the RS case.

\vspace{5mm}

In conclusion, we have studied the effects of the replica symmetry breaking
in the renormalization group for the 2D random Ising and Baxter models.
We have found that the traditional replica-symmetric renormalization group
flows are stable with respect to the RSB modes. Thus, unlike the other
suggestions \cite{ziegler}, we conclude that
the effects of possible non-perturbative degrees of freedom in the 2D random
Ising and Baxter models must be irrelevant for the critical behaviour studied
earlier.


\vspace{1cm}

\begin{thebibliography}{99}

\bibitem{harris} A.B.Harris, J.Phys. {\bf C 7}, 1671 (1974)

\bibitem{new}
T. C. Lubensky and A. B. Harris, AIP Conf. Proc. {\bf 24} 311 (1974)\\
G. Grinstein, AIP Conf. Proc. {\bf 24} 313 (1974)\\
A. B. Harris and T. C. Lubensky, Phys.Rev.Lett., {\bf 33}, 1540 (1974)\\
D. E. Khmelnitskii, ZhETF (Soviet Phys. JETP) {\bf 68}, 1960 (1975)\\
G. Grinstein and A. Luther, Phys.Rev. {\bf B 13}, 1329 (1976)

\bibitem{ising}
Vik.S.Dotsenko and Vl.S.Dotsenko, JETP Lett., {\bf 33}, 37 (1981)\\
Vik.S.Dotsenko and Vl.S.Dotsenko, Adv.Phys. {\bf 32}, 129 (1983)\\
B.N.Shalaev, Sov. Phys. Solid State, {\bf 26}, 1811 (1984)\\
R.Shankar, Phys.Rev.Lett., {\bf 58}, 2466 (1986)\\
A.W.W.Ludwig, Phys.Rev.Lett., {\bf 61}, 2388 (1988)

\bibitem{baxter}
Vik.S.Dotsenko and Vl.S.Dotsenko, J. Phys. A: Math.Gen. {\bf 17}, L301 (1984)

\bibitem{potts}
A.W.W.Ludwig, Nucl.Phys. B {\bf 285}, 97 (1987);
Nucl.Phys. B {\bf 330}, 639 (1990)\\
Vl.S.Dotsenko, M.Picco and P.Pujol, Phys.Lett. B {\bf 347}, 113 (1995)\\
Vl.S.Dotsenko, M.Picco and P.Pujol, Nucl.Phys. B {\bf 455}, 701 (1995)

\bibitem{numer}
V.B.Andreichenko, Vl.S.Dotsenko, W.Zelke and J.-S.Wang,
Nucl.Phys. B {\bf 344}, 531 (1990)\\
J.-S.Wang, W.Zelke,  Vl.S.Dotsenko and V.B.Andreichenko,
Europhys.Lett., {\bf 11}, 301 (1990)\\
J.-S.Wang, W.Zelke,  Vl.S.Dotsenko and V.B.Andreichenko,
Physica A., {\bf 164}, 221 (1990)\\
W.Zelke, L.N.Shchur and A.L.Talapov, in "Annual Reviews of Computational
Physica" (Ed. D.Stauffer), World Scientific, Singapore, {\bf 17} (1994)

\bibitem{sg}
M.Mezard, G.Parisi, M.Virasoro, "Spin glass theory and beyond",
World Scientific, 1987

\bibitem{manifold}
M.Mezard and G.Parisi, J.Phys. I {\bf 1}, 809 (1991)

\bibitem{rsb-Marc}
M.Mezard and A.P.Young, Europhys.Lett., {\bf 18}, 653 (1992)

\bibitem{Korshunov}
S. Korshunov, Phys.Rev. {\bf B48}, 3969 (1993)

\bibitem{dhss}
Vik.S.Dotsenko, B.Harris, D.Sherrington and R.Stinchbombe,
J.Phys. A: Math.Gen. {\bf 28}, 3093 (1995)

\bibitem{df}
Vik.S.Dotsenko and D.E.Feldman, J.Phys.A {\bf 28}, 5183 (1995)

\bibitem{doussal}
P. Le Doussal and T.Giamarchi, Phys.Rev.Lett., {\bf 74}, 606 (1995)

\bibitem{rsb-potts}
Vik.S.Dotsenko, Vl.S.Dotsenko, M.Picco and P.Pujol,
Europhys.Lett., {\bf 32}, 425 (1995)

\bibitem{HG}
C.A.Hurst and H.S.Green, J.Chem.Phys. {\bf 33}, 1059 (1960)

\bibitem{B} R.J.Baxter, "Exactly solved models in statistical mechanics",
Academic Press, 1982

\bibitem{ziegler} K.Ziegler, Nucl.Phys. B {\bf 285}, 606 (1987)


\end{thebibliography}
\end{document}